# The Nexus of Science Fiction, Box Office Success, and Technology Representation: A Case Study of the Marvel Cinematic Universe

Iqra Tariq

**Abstract**— This paper investigated the applied science domains and subjects depicted in Marvel Cinematic Universe (MCU) movies and assessed the relationship between technological portrayal and box office success. The study looked at 164 publications in academic literature that employed MCU movies. In addition to the foregoing, the study discovered that MCU movies have been used in academic literature in a variety of ways, including teaching science ideas, analyzing ethical dilemmas, and examining social and cultural themes. This shows that MCU movies could be used for educational and societal goals as well. Also, the study demonstrates that MCU movies are more than just popular entertainment. They can also be used to teach science, investigate ethical dilemmas, and investigate social and cultural themes. According to our main investigation, MCU movies represent a wide range of applied science topics, such as technology, magic, ancient technology, cosmic technology, multiverse technology, energy, physics, and engineering. Also, the portrayal of technology is important in the popularity of Science Fiction movies because audiences are drawn to the spectacle of sophisticated technology and the spectacular action sequences that it allows. The study also discovered that the depiction of technology is associated with box office performance, with movies with a higher *Tech Content Count* being more profitable.

**Keywords:** Film Success Factor, Machine Learning (ML), Data Analytics, Technology Representation.

---

**Note:** This paper is an empirical case study examining the correlation between the density of computational technology representation and box office performance. Building on a preliminary analysis of the first four phases of the Marvel Cinematic Universe (MCU), which demonstrated a significant "tech boost" effect on commercial success, this research expands the scope to a broader sci-fi dataset to validate these trends across different productions and larger dataset.

# 1. Introduction

With the release of "Iron Man" in 2008, MCU was launched, and it has since established itself as a groundbreaking and successful movie series [1]. Audiences first met enduring figures like Iron Man, Captain America, and the Avengers in this interconnected universe. The MCU stands apart due to its dedication to maintaining narrative coherence throughout all its movies and television shows, which encourages fan interaction and a sense of community [2]. MCU movies have regularly performed well financially and set standards for industry. In addition to its commercial success, the MCU has been instrumental in advancing diversity and representation in the superhero genre and tackling difficult societal topics [3]. It is the perfect subject for examining technology representation and Box Office success because of its wide range of genres, subjects, and personalities that have profoundly influenced contemporary cinema [4].

Science Fiction movies that feature technology are extremely important because they act as a prism through which society considers the ramifications of innovative technologies [5]. These movies, which frequently show both utopian and apocalyptic futures, promote conversations about moral, sociological, and ethical issues [6]. They also investigate fundamental facets of human nature, stimulating creativity while expressing concern about unrestrained technological progress. Additionally, audiences are enthralled by Science Fiction's visual medium, which provides amusement and inspiration, and legendary movies have had a lasting cultural impact [7]. Fundamentally, the depiction of technology in Science Fiction serves as a potent weapon for igniting the imagination, influencing technological discourse, and inspiring real-world innovation [5-8].

## 1.1  Aim and Objectives

This research paper's main aim is to investigate the connection between Science Fiction movies' Box Office performance and how technology is portrayed in those movies.

The following research objectives will be addressed to reach the aim:

1. List and group the applied science fields that the MCU movies depict.
2. Evaluate how much the portrayal of technology contributes to the Box Office success of MCU movies.

## 1.2  Significance of the Study

The global Science Fiction movie market is predicted to reach $18.2 billion in 2023, making it the most popular genre of movies, accounting for approximately 15% of all Box Office revenue [29]. This study's findings will help understand the relationship between technology representation and the Box Office performance of Science Fiction movies. The information presented in the paper will be helpful to moviemakers, producers, and other participants in the movie industry providing insight into technology importance and representation style.

The investigation into the MCU movies will utilize a case study methodology. The MCU is a well-known and prosperous Science Fiction series that uses a variety of technology, making it an excellent case study for this investigation.

## 2.  Literature Review

Le Voyage dans la Lune (A Trip to the Moon), directed by Georges Méliès, was released in 1902, marking the beginning of Science Fiction cinema. This creative movie, which employed special effects to portray a trip to the moon, was a smash hit with viewers [9,14]. Science Fiction movies gained popularity throughout the ensuing decades (1900s-1940s) and expanded their subject matter. Things to Come (1936), and Metropolis (1927) are a few of the best Science Fiction movies from this time. These movies explored issues including artificial intelligence, space travel, and alien life [10,14]. The 1950s and 1960s were a golden age for Science Fiction cinema, with movies such as 2001: A Space Odyssey (1968), and Planet of the Apes (1968) pushing the boundaries of the genre [10,14]. Complex topics including the nature of humankind, the hazards of technology, and the value of individual freedom were examined in these movies. [11]. In the 1970s and 1980s, Science Fiction movies became more mainstream and began to incorporate elements of other genres, such as action, adventure, and comedy [12,14]. Some of the most popular Science Fiction movies of this era include A Clockwork Orange (1971), Star Wars (1977), Alien (1979), and Back to the Future (1985). These movies were well-liked by a large audience and contributed to the growth of the Science Fiction genre. [10,12].

Science Fiction movies have developed and explored new themes and ideas recently. [13]. Some of the most notable Science Fiction movies of the 21st century include The Matrix (1999), Children of Men (2006), Arrival (2016), Everything Everywhere All at Once (2022), and Mission: Impossible 7 (2023). These movies have received recognition for their intricate storylines, innovative spectacular effects, and challenging subjects [7,10,14]. Table 1 presents the Science Fiction Movies Eras with respect to popular themes and notable movies.

Table 1: Science Fiction Movies Eras and Respective Trending themes [14,15]

| Era | Notable movies | Themes and trends |
|---|---|---|
| Early cinema (1902-1940s) | Le Voyage dans la Lune (1902), Metropolis (1927), Things to Come (1936) | Technological advancement, artificial intelligence, alien life |
| Golden age of Science Fiction (1950s-1960s) | (1968), Planet of the Apes (1968), A Clockwork Orange (1971), Star Trek (1966-1969) | Space exploration, dystopia, social commentary |
| New Hollywood Era (1970s-1980s) | Star Wars (1977), Alien (1979), Back to the Future (1985), Blade Runner (1982) | Blockbuster action movies, special effects, social commentary |
| Modern Era (1990s-present) | The Matrix (1999), Arrival (2016), Interstellar (2014), Blade Runner 2049 (2017), Dune (2021), Everything Everywhere All at Once (2022) | Diversity and inclusion, climate change, social inequality, the future of technology |

Among big banners in Science Fiction movies' production, MCU has left an indelible mark on the movie industry, achieving unparalleled success both financially and culturally [16]. Notably, "Avengers: Endgame" (2019) stands as the highest-grossing movie of all time, surpassing "Avatar," with a global Box Office gross exceeding $2.7 billion [17-19] on their first release. This financial achievement has set a new standard for blockbuster movies. What distinguishes the MCU is its pioneering shared cinematic universe model, where characters and narratives seamlessly cross over between individual movies and television series [20]. This interconnected storytelling approach has not only captivated audiences but has also influenced other studios to emulate this successful formula. The MCU's commitment to diversity and inclusion is evident in movies like "Black Panther" (2018) and "Captain Marvel" (2019), resonating with global audiences and promoting underrepresented voices [21].

Furthermore, the MCU's genre-blending, combining superhero narratives with elements of space opera, political thriller, and fantasy, has broadened the appeal of the superhero genre. Beyond cinema, the MCU has successfully expanded into television and streaming platforms, with series such as "WandaVision" and "The Falcon and the Winter Soldier" contributing to its narrative scope [22]. This transition aligns with the wider industry trend of content distribution through streaming services like Disney+. MCU's cultural impact is exemplified by its dedicated fan base, merchandise sales, and conventions, creating a robust fan culture that extends beyond the screen. The influence of the MCU extends to how studios approach franchise building, emphasizing long-term planning, interconnected storytelling, and character development [23]. It has become a trailblazer in contemporary moviemaking, reshaping industry trends, and demonstrating the enduring significance of cinematic universes in the world of cinema [21-23].

Some of the key achievements of MCU are 1: Highest-grossing movie franchise of all time with 29.55 billion US dollars [24], 2: Most critically acclaimed cinematic universe of all time with 26 Academy Award nominations and 4 wins [25], 3: Most popular shared universe of movies 30 movies released and 10 television series released, with more to come [26], The MCU has inspired other studios to create their own shared universes, such as the DC Extended

Universe and the Warner Bros. Discovery cinematic universe [27], 4: Most diverse movie franchise of all time with 57% of the cast and crew of the MCU movies are women, people of color, and all genders [28].

Not only is this MCU widely used in scholarly articles for different areas of research. MCU has been used in a variety of domains for the research purposes, including Computing: To study AI, machine learning, and other advanced technologies, Economics: To study economic inequality, sustainable development, and the global economy, Arts, and humanities: To study identity, power, colonialism, and social justice, Science: To study physics, chemistry, and biology, social sciences: To study psychology, sociology, and political science. MCU movies have also been used in literature to explore ethical, philosophical, and cultural themes [30].

To study the role of MCU in literature; Kaggle dataset "*Marvel Cinematic Universe in Literature Till 9/23*" [30] is used to investigate the domains and respective applications areas. The dataset contains 169 entries where 5 entries are excluded due to being the book reviews then further data is preprocessed for the check of sub domains and application areas. Table 2 shows the final research items which are studied from the dataset.

Table 2 Final Investigation Items related to MCU from Dataset

| Publishers | Papers | Publishers | Papers | Publishers | Papers |
|---|---|---|---|---|---|
| IEEE | 1 | PubMed | 4 | ScienceDirect | 13 |
| MDPI | 6 | Tylor And Francis | 96 | ACM Digital Library | 9 |
| Springer Link | 35 | **Total** | **164** | | |

The Dataset exploration shows that there are 9 main knowledge domains in which MCU literature papers are published. Figure 1 maps the number of articles related to each domain, showing *Arts and Humanity* is the most used domain for the study of MCU movies followed by *Computing* and *Science*.

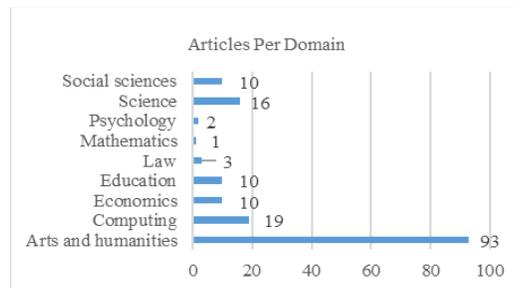

Figure 1 Articles' Domain Relevance

Not only this, but each domain is also further studied with respect to sub domains targeting collective 169 application areas, Providing its diverse significance for research perspective. Figure 2 shows the applications areas associated with each main domain of the dataset.

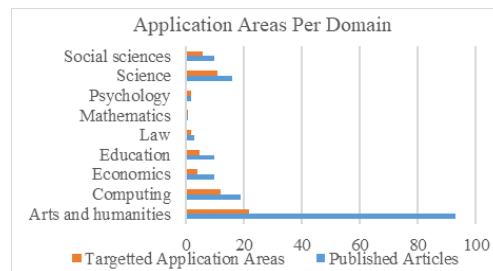

Figure 2 Application Areas with respect to Each Domain

## 2.1 Gaps in Existing Literature

Even though MCU is employed in 169 research items from seven big publishers [30], there is no single research work available on the role of MCU in applied sciences portrayal and the role of technology in the success of any Science Fiction movie. Most studies concentrated on the MCU in relation to a specific topic, such as Science, Arts & Humanities, and Mathematics. Furthermore, it has not been determined how much the pace of technological portrayal has affected the movie Box Office performance in the MCU, which is related to the response on financial success.

## 3. Methodology

To achieve the desired investigation and research outcomes we have distributed the entire process in five phases. Figure 3 shows the main methodology of the paper showing five main phases and their respective sub activities we are intended to perform during the research process.

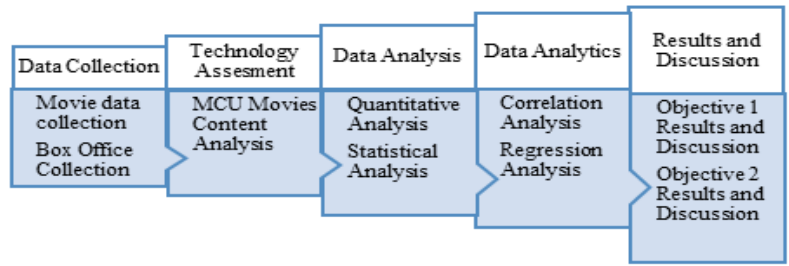

Figure 3 Methodology of the Research Work

Moreover, section IV elaborates each activity and every method, material, and tools which we have used for the desired outcome.

## 4. Material, Methods, and Tools
### 4.1 Data Collection

#### 4.1.1 Selection of MCU Movies for Analysis

The focus of our study lay on MCU movies, for that purpose we have collected the list of MCU movies from Marvel official website [17] and used Numbers [18] for the finalization and confirmation of respective release date. The study only focuses on the first four phases of MCU release schedule which covers 14 years from Iron Man released in 2008 to Black Panther: Wakanda Forever in 2022. Figure 4. Displays all thirty movies' official posters.

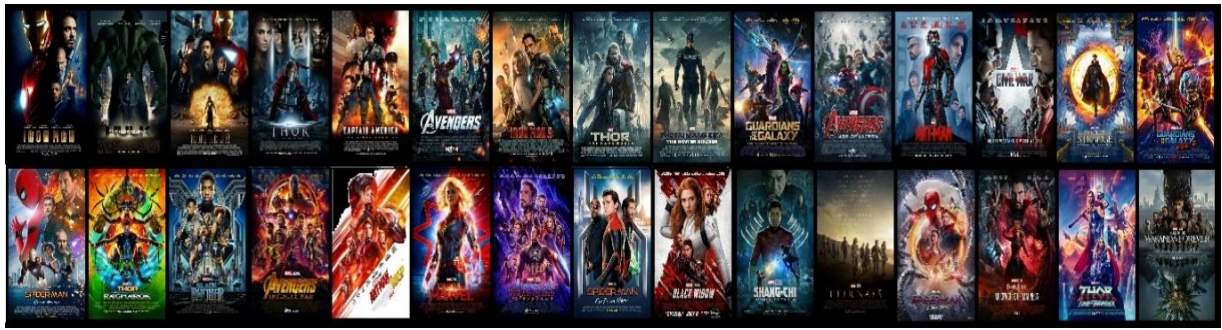

Figure 4 MCU's Phase I-IV All Movies Posters

Table 3. represents the details of thirty movies which have been used for the main study, all movies are assigned code with respect to the movie sequence.

Table 1 MCU Movies and Release Date

| No. | Movies | Release Date | No. | Movies | Release Date |
|---|---|---|---|---|---|
| MCU-M1 | Iron Man | 5/2/08 | MCU-M16 | Spider-Man: Homecoming | 7/7/17 |
| MCU-M2 | The Incredible Hulk | 6/13/08 | MCU-M17 | Thor: Ragnarok | 11/3/17 |
| MCU-M3 | Iron Man 2 | 5/7/10 | MCU-M18 | Black Panther | 2/16/18 |
| MCU-M4 | Thor | 5/6/11 | MCU-M19 | Avengers: Infinity War | 4/27/18 |
| MCU-M5 | Captain America: The First Avenger | 7/22/11 | MCU-M20 | Ant-Man and the Wasp | 7/6/18 |
| MCU-M6 | The Avengers | 5/4/12 | MCU-M21 | Captain Marvel | 3/8/19 |
| MCU-M7 | Iron Man 3 | 5/3/13 | MCU-M22 | Avengers: Endgame | 4/26/19 |
| MCU-M8 | Thor: The Dark World | 11/8/13 | MCU-M23 | Spider-Man: Far from Home | 7/2/19 |
| MCU-M9 | Captain America: The Winter Soldier | 4/4/14 | MCU-M24 | Black Widow | 7/9/21 |
| MCU-M10 | Guardians of the Galaxy | 8/1/14 | MCU-M25 | Shang-Chi and the Legend of the Ten Rings | 9/3/21 |
| MCU-M11 | Avengers: Age of Ultron | 5/1/15 | MCU-M26 | Eternals | 11/5/21 |
| MCU-M12 | Ant-Man | 7/17/15 | MCU-M27 | Spider-Man: No Way Home | 12/17/21 |
| MCU-M13 | Captain America: Civil War | 5/6/16 | MCU-M28 | Doctor Strange in the Multiverse of Madness | 5/6/22 |
| MCU-M14 | Doctor Strange | 11/4/16 | MCU-M29 | Thor: Love and Thunder | 7/8/22 |
| MCU-M15 | Guardians of the Galaxy Vol. 2 | 5/5/17 | MCU-M30 | Black Panther: Wakanda Forever | 11/11/22 |

#### 4.1.2 Box Office Data Collection

One of the main focuses of this study is to find the correlation of Box Office success and technology representation, so the Box Office facts were carefully collected from three main movie websites, IMDB [10], Numbers [18] and Box Office Mojo [19] and verified for the final confirmation. The Box Office data of each movie is presented in Table 6.

### 4.2 Data Analysis Methods

#### 4.2.1 Technology Representation Analysis

We have performed content analysis of MCU movies to identify the applied sciences domains that are portrayed and to assess the extent to which technology is portrayed to fulfil our research objectives. We have also gathered and analyzed data on the Box Office performance of MCU films.

For content analysis of thirty MCU movies we have identified three main elements to discover from the movies Following are the main content analysis queries.

1. Domain of applied Sciences which is portrayed in the movie.
2. Sector targeted using the domain application.
3. artifacts used related to that domain covering the sector of study.

Table 4. shows the data of content analysis which was discovered from thirty movies.

Table 2 Content Analysis Data from MCU Movies

| Movies | Artefact | Sector | Domain |
|---|---|---|---|
| MCU-M1 | Mark I - VII armor, JARVIS AI | Artificial Intelligence (AI), robotics, exoskeletons | AI, robotics, physics, and engineering |
| MCU-M2 | Emil Blonsky's transformation into Abomination, Bruce Banner's research on gamma radiation | Gamma radiation, bioengineering | Biochemistry, genetics, and engineering |
| MCU-M3 | Mark VI armor, Ivan Vanko's whips, S.H.I.E.L.D. technology | AI, robotics, energy | AI, robotics, and physics |
| MCU-M4 | Thor's hammer Mjölnir, Bifröst bridge | Asgardian technology | Physics, engineering, and magic |
| MCU-M5 | Super Soldier Serum, Howard Stark's inventions | Super Soldier Serum, Stark technology | Biochemistry, genetics, and engineering |
| MCU-M6 | Helicarrier, Iron Man's armor, Tesseract, JARVIS AI | Helicarrier, Stark technology, Tesseract, and armor | Engineering, physics, and energy, AI, and robotics |
| MCU-M7 | Mark VIII - XLII armor, Extremis virus, JARVIS AI | AI, robotics, Extremis virus | AI, robotics, physics, and biochemistry |
| MCU-M8 | Aether, Reality Stone | Aether, Reality Stone | Physics, energy, and magic |
| MCU-M9 | Project Insight, helicarriers, Hydra technology, FRIDAY | Project Insight, helicarriers, Hydra technology | Engineering, physics, and AI |
| MCU-M10 | Milano spacecraft, Orb, Power Stone | Milano spacecraft, Orb, Power Stone | Engineering, physics, and energy |
| MCU-M11 | Ultron, Vision | Ultron, Vision | AI, robotics, energy, and physics |
| MCU-M12 | Pym particles, Ant-Man suit | Pym particles, Ant-Man suit | Physics, engineering, AI, and material science |
| MCU-M13 | Iron Man's armor, Captain America's shield, Vision | Iron Man's armor, Captain America's shield, Vision | Engineering, physics, AI, and materials science |
| MCU-M14 | Eye of Agamotto, Time Stone, Sling Ring portals | Eye of Agamotto, Time Stone, Sling Ring portals | Physics, energy, and magic |
| MCU-M15 | Ego the Living Planet, Adam Warlock's cocoon | Ego the Living Planet, Adam Warlock's cocoon | Physics, biology, and engineering |
| MCU-M16 | Spider-Man suit, Vulture's wings | Spider-Man suit, Vulture's wings | Engineering, physics, AI, and materials science |
| MCU-M17 | Hulkbuster armor, Sakaarian technology | Hulkbuster armor, Sakaarian technology | Engineering, physics, and materials science |
| MCU-M18 | Vibranium, Wakandan technology, Shuri's lab | Vibranium, Wakandan technology | Engineering, physics, AI, and materials science |
| MCU-M19 | Infinity Gauntlet, Infinity Stones, Vision | Infinity Gauntlet, Infinity Stones | Physics, energy, AI, and magic |
| MCU-M20 | Quantum Realm, Wasp suit | Quantum Realm, Wasp suit | Physics, engineering, AI, and quantum physics |
| MCU-M21 | Binary fission powers, light speed travel | Binary fission powers, light speed travel | Physics, energy, and space travel |
| MCU-M22 | Time travel, Quantum Realm, nanotech armor | Time travel, Quantum Realm, nanotech armor | Physics, quantum physics, AI, and space travel, Robotics |

| MCU-M23 | E.D.I.T.H., Mysterio's illusions | E.D.I.T.H., Mysterio's illusions | Engineering, physics, AI, and Robotics |
|---|---|---|---|
| MCU-M24 | Red Room technology, Taskmaster's mimicry chip | Red Room technology, Taskmaster's mimicry chip | Engineering, physics, and bioengineering |
| MCU-M25 | Ten Rings, Ta Lo | Ten Rings, Ta Lo | Physics, magic, and ancient technology |
| MCU-M26 | Celestial technology, Eternals' powers | Celestial technology, Eternals' powers | Physics, energy, and cosmic technology |
| MCU-M27 | Multiverse portal, Green Goblin's pumpkin bombs, Iron Spider suit's AI | Multiverse portal, Green Goblin's pumpkin bombs, Iron Spider suit's AI | Physics, engineering, and multiverse technology, AI, and Robotics |
| MCU-M28 | Darkhold, America Chavez's powers | Darkhold, America Chavez's powers | Magic, physics, and multiverse technology |
| MCU-M29 | Stormbreaker, Zeus' Lightning Bolt | Stormbreaker, Zeus' Lightning Bolt | Physics, magic, and ancient technology |
| MCU-M30 | Vibranium, Wakandan technology, Shuri's lab | Vibranium, Wakandan technology | Engineering, physics, AI, and materials science |

### 4.2.2  Domain and Technology Representation Analysis

Once the content analysis has been finalized and verified, further all the domains are classified with respect to the main applied science area resulting in the four super domains (Technology, Science, Natural Science and Fictional Technology).

Table 3 List of Scientific Application Areas/Domains Targeted by MCU Movies with No. of Occurrences

| Domains | Occurrences | Domains | Occurrences |
|---|---|---|---|
| **Technology** | | **Science** | |
| AI | 15 | Biochemistry | 3 |
| Robotics | 8 | Biology | 1 |
| **Fictional Technology** | | Genetics | 2 |
| Magic | 7 | Bioengineering | 1 |
| Ancient Technology | 2 | **Natural Science** | |
| Cosmic Technology | 1 | Materials Science | 6 |
| Multiverse Technology | 1 | Quantum Physics | 2 |
| Energy | 8 | Engineering | 18 |
| | | Physics | 28 |

### 4.2.3  Statistical Analysis of Box Office Performance

We have used a descriptive statistical analysis approach for the quantitative analysis of the Box Office performance [39]. To identify the profit percentage, we have used two major features of each movie *Budget* and *Box Office*. Using these two features we have calculated the *Profit* of each movie. Further Equation 1 and 2 have been used to calculate one of the final analysis features *Profit Percentage* based on the *Profit* attribute.

$$Profit = Box\ Office - Budget \quad (1)$$

$$Profit\ Percentage = \left(\frac{Profit}{Budget}\right) * 100 \quad (2)$$

9Table 4 Statistical analysis of Box Office performance

| Movie | Budget | Box Office | Profit | Profit Percentage | Tech Content Count | Class |
|---|---|---|---|---|---|---|
| MCU-M2 | 150 | 264.8 | 114.8 | 76.53333333 | 0 | 4 |
| MCU-M24 | 200 | 379.7 | 179.7 | 89.85 | 0 | 4 |
| MCU-M26 | 200 | 402 | 202 | 101 | 0 | 4 |
| MCU-M5 | 140 | 370.4 | 230.4 | 164.5714286 | 0 | 4 |
| MCU-M25 | 150 | 432.2 | 282.2 | 188.1333333 | 0 | 4 |
| MCU-M4 | 150 | 449.3 | 299.3 | 199.5333333 | 0 | 4 |
| MCU-M3 | 200 | 623.9 | 423.9 | 211.95 | 2 | 4 |
| MCU-M8 | 175 | 644.8 | 469.8 | 268.4571429 | 0 | 4 |
| MCU-M29 | 200 | 760.9 | 560.9 | 280.45 | 0 | 3 |
| MCU-M20 | 162 | 622.7 | 460.7 | 284.382716 | 1 | 3 |
| MCU-M12 | 130 | 519.3 | 389.3 | 299.4615385 | 1 | 3 |
| MCU-M14 | 165 | 677.7 | 512.7 | 310.7272727 | 0 | 3 |
| MCU-M1 | 140 | 585.1 | 445.1 | 317.9285714 | 2 | 3 |
| MCU-M9 | 170 | 714.4 | 544.4 | 320.2352941 | 1 | 3 |
| MCU-M30 | 200 | 842.4 | 642.4 | 321.2 | 1 | 3 |
| MCU-M15 | 200 | 863.7 | 663.7 | 331.85 | 0 | 3 |
| MCU-M13 | 250 | 1,100 | 850 | 340 | 1 | 3 |
| MCU-M10 | 170 | 773.3 | 603.3 | 354.8823529 | 0 | 3 |
| MCU-M17 | 180 | 853.9 | 673.9 | 374.3888889 | 0 | 3 |
| MCU-M28 | 200 | 955.8 | 755.8 | 377.9 | 0 | 3 |
| MCU-M16 | 175 | 880.2 | 705.2 | 402.9714286 | 1 | 3 |
| MCU-M11 | 250 | 1,400 | 1150 | 460 | 2 | 3 |
| MCU-M7 | 200 | 1,200 | 1000 | 500 | 2 | 2 |
| MCU-M21 | 175 | 1,100 | 925 | 528.5714286 | 0 | 2 |
| MCU-M18 | 200 | 1,300 | 1100 | 550 | 1 | 2 |
| MCU-M19 | 300 | 2,000 | 1700 | 566.6666667 | 1 | 2 |
| MCU-M6 | 220 | 1,500 | 1280 | 581.8181818 | 2 | 1 |
| MCU-M23 | 160 | 1,100 | 940 | 587.5 | 2 | 1 |
| MCU-M22 | 356 | 2,700 | 2344 | 658.4269663 | 2 | 1 |
| MCU-M27 | 200 | 1,900 | 1700 | 850 | 2 | 1 |

#### 4.2.4 Data Analytics

For finding the insight using data analytics based on movie Box Office and technology nexus we have used Correlation analysis and Regression analysis method. Firstly, each movie is assigned the score from 0 to 2 as per the occurrence of technology domain sector from Table 6. Secondly, we have used histogram method using MS. Excel to find the four equal bins to distribute the profit percentage in four classes (Ranges from 76.53 to 850), where 1 represents the most profitable and 4 represents the lowest profitable movie.

$$= \text{HISTOGRAM}(A2:A31, \{76.53, 276.533, 476.533, 676.533, 876.533\}) \quad (3)$$

Histogram Function Equation 3 is used to identify the range of classes from Profit Percentage feature. The main reason to identify the Profit Percentage feature was to find Correlation Coefficient and performing regression analysis on MCU movies.

Figure 5 shows the range and occurrence of each bin based on profit percentage.

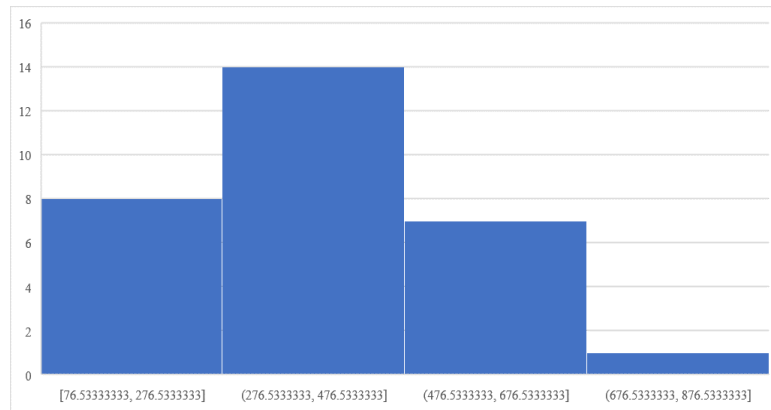

Figure 5 Profit Percentage Class Distribution Bins

Further equation 4 is used to identify the class against each movie, so that data analytics are performed. Table 6 shows the final features *Profit Percentage* and *Class*.

$$Class = \begin{cases} 4 \text{ if } 76.53 < Profit\ Percentage\ < 276.53 \\ 3 \text{ if } 276.53 < Profit\ Percentage\ < 476.53 \\ 2 \text{ if } 476.53 < Profit\ Percentage\ < 676.53 \\ 1 \text{ if } 676.53 < Profit\ Percentage\ < 876.53 \end{cases} \quad (4)$$

To get the analytics from the movie data we have used five regression algorithms of machine learning. (Linear Regression, Simple Linear Regression, Multilayer Perceptron, Random Forest, and Random Tree). The reason for the selection of these algorithms was the nature of our objectives, as it inquires the relativity and dependency of technology and Box Office [35-38]. Further Correlation Coefficient is used to identify the relationship between technology representation and Box office success. The intended new features were *Tech Content Count* representing the technology count and *Class* representing the Box Office performance level. The data is presented in Table 6. After this we performed regression analysis to get the values of four main quantitative performance measures to find the error rate from the data (Relative Absolute Error (RAE), and Root Relative Square Error (RRSE), Mean Absolute Error (MAE), Root Mean Square Error (RMSE)).

### 4.3 Tools

We have used two main applications for our study, first we have used MS. Excel for descriptive and quantitative analysis, and lastly, we have used Weka for data analytics to calculate research objectives' outcomes. Moreover Google Bard.ai, Grammarly and Chat GPT have been used for the final grammar check and spelling check and confirmation.

## 5. Technology Representation in the Marvel Cinematic Universe

Table 6 shows the list of all Super Domains, Domains, and their Occurrences in the MCU movies. Dominated by Physics, Engineering and Artificial Intelligence, the collective count shows the prominent contribution of technology and natural sciences in the MCU movies.

### 5.1 Overview of Technology Themes in MCU Movies

The super domain of the investigation is Technology, from table 6. We have identified that total domain occurrence is 23, further distributed in AI and Robotics with the rate of 15 and 8.

### 5.2 Key trends and patterns in technology representation from MCU

The following are important themes and patterns in technology representation which we have derived from MCU movie content analysis data from Table 4.

1. The main topic is artificial intelligence (AI). It is likely that AI is used in other technologies as well, such as the Spacecraft and The Arc Reactor. This proves that AI is a crucial driver of future technological advancement and representation in the selected movie dataset.
2. Another significant trend is robotics. Eight of the technologies listed mention robotics, and it is likely that robotics is used in other technologies as well, such as AI Armor and The Arc Reactor. This shows that robots are playing a larger role in technology projection.
3. As we know Space exploration is getting increasingly accessible. According to the spacecraft technologies identified from content analysis, space travel is also fascinating and stimulating projection This has opened new avenues for space exploration and colonization in the masses.
4. The integration of technologies is widely used in these movies. Some of the listed technologies, such as AI Armor and E.D.I.T.H., are interconnected.

### 5.3 Examining the portrayal of technology in MCU.

AI and Robotics are not only directly used in the movies but also the applications of these two domains are identified from the content analysis data from Table 4. These domains have used interconnectivity portraying smart devices and artifacts on the screen. Helicarrier (Transportation), Armor (Security and Defense) and AI suits (Wearables) are some of the main sectors targeted by the technology in the data. Table 8 provides all the technology-oriented sectors from our study which is derived from content analysis data from Table 5.

Table 5 All Technologies Represented in MCU Movies

| Technology Sectors | | | |
|---|---|---|---|
| AI | Stark Technology | AI Armor | Exoskeletons |
| Robotics | Quantum Realm | Spacecraft | The Arc Reactor |
| Helicarrier | Digital Illusions | E.D.I.T.H. | AI Suit |

Each artifact identified from the data shows an advanced level processing and application in the respective movie. This portrayal of technology has inspired a prominent range of domain learning and innovation.

Main technology artifacts from MCU movies are presented in Table 8.

Table 6 All Computing Technology Artefacts s in MCU Movies

| Technology Artefacts | | | |
|---|---|---|---|
| Mark I - XLII armor | Vision | Vulture's wings | S.H.I.E.L.D. technology |
| JARVIS AI | Ant-Man suit | Hulkbuster armor | Stark technology arc reactor |
| Mark VI armor | Iron Man's armor | Quantum Realm | Helicarrier |
| Ivan Vanko's whips | Spider-Man suit | Wasp suit | FRIDAY AI |
| Ultron | Mysterio's illusions | E.D.I.T.H. | Milano spacecraft |

## 6. Box Office Success in the Marvel Cinematic Universe
### 6.1 Box Office performance of MCU films

The Box Office results for the MCU have increased steadily over time. An MCU movie's average box office take has climbed from $399 million in Phase One (2008-2012) to $723 million in Phase Two (2013-2015), $873 million in Phase Three (2016-2019), and $956 million in Phase Four (2021-2022). MCU have a promising rate when it comes to the Box Office sector. Table 7 shows that not one of any movies of MCU have gone under loss. On other side the data also shows that there is a diverse trend in the profit percentage making some movies more successful than others. Table 6. shows the list of all movies' Box Office calculated profit percentage in ascending order. The list ends with most profitable movie for MCU is *Spiderman: No Way Home* started by *The Incredible Hulk* making this movie the lowest profitable on Box Office.

### 6.2 Role of technology representation in attracting audiences

The portrayal of technology is crucial for drawing viewers [32]. It can be used to make experiences that are more immersive and interesting, to represent other voices and viewpoints, and to tell stories that would not be feasible otherwise [31-33]. With the aid of special effects, three generations of Spider-Man actors could appear in "Spider-Man: No Way Home" (2021). With $1.9 billion in global box office receipts, the movie was a major hit [18]. This shows the acceptance of the content among the audience which is observed by the Box Office collection of these movies. Overall, the secret to unleashing the full potential of storytelling and the future of entertainment is technology representation [34].

## 7. Relationship between Technology Representation and Box Office Success

### 7.1 Statistical Analysis Results

We have created a pivot table considering two derived features from table 7, *Tech Content Count* and *Class*. Table 9 shows the occurrence of each Class with respect to relative Tech Content Count.

Table 7 Tech Content Count Vs. Class

| Tech Content Count | Class | | | |
|---|---|---|---|---|
| | 1 | 2 | 3 | 4 |
| 0 | | 1 | 6 | 7 |
| 1 | | | 2 | 6 |
| 2 | 4 | 1 | 2 | 1 |

The following Figure 6. visually represents that no movie falls in class 1 if the Tech Content Count is 0, similarly all highest profitable movies belong to class 1. Also, a substantial number of movies (7) fall in class 4 where tech content count is 0.

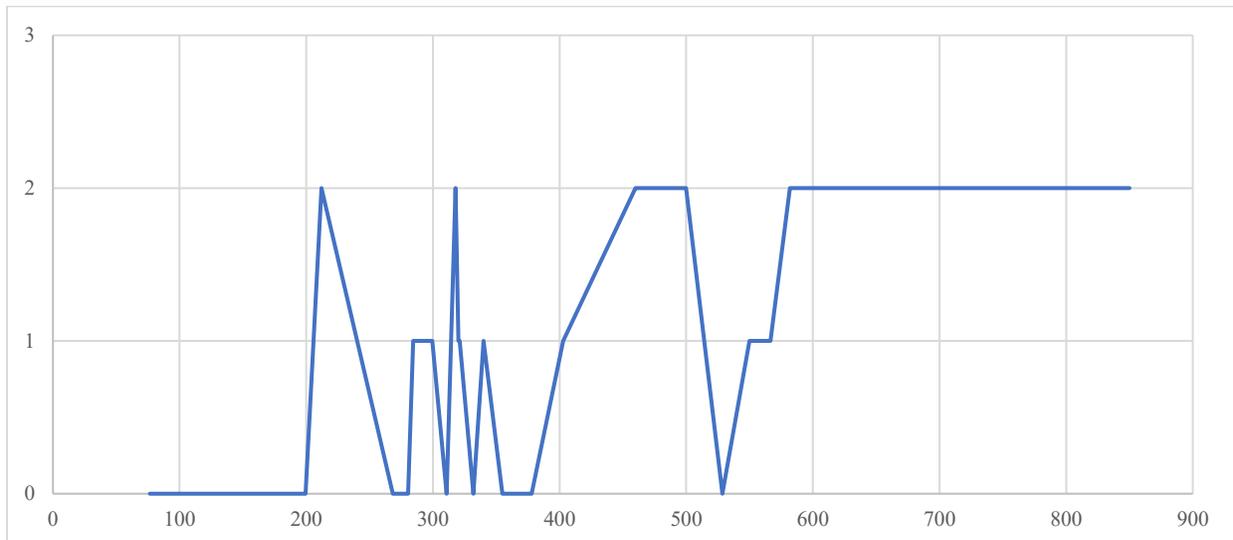

Figure 6 Tech Content Count Representation

## 7.2 Data Analytics Results

### 7.2.1 Correlation Analysis

The results of correlation coefficient between the final features (Tech Content count, Class, Profit Percentage, and Index) of the study are presented in Table 10. Among all the regression algorithms Random Forest has shown the best results as compared to the rest of the five algorithms.

Table 10 Correlation Coefficients Results

| Index | Algorithm | Correlation Coefficient |
|---|---|---|
| 1 | Linear Regression | 0.9078 |
| 2 | Simple Linear Regression | 0.9211 |
| 3 | Multilayer Perceptron | 0.8824 |
| 4 | *Random Forest* | *0.964* |
| 5 | Random Tree | 0.953 |

### 7.2.2 Regression Analysis

The results of quantitative performance measures between the final features (Tech Content count, Class, Profit Percentage, and Index) of the study are presented in Table 11. Among all the regression algorithms Random Forest has shown the least error margin as compared to the rest of the five algorithms.

Table 8 Rate of Errors from the Data

| Index | Algorithm | MAE | RMSE | RAE | RRSE |
|---|---|---|---|---|---|
| 1 | Linear Regression | 0.3114 | 0.4059 | 41.62 | 41.46 |
| 2 | Simple Linear Regression | 0.2888 | 0.3737 | 38.60 | 38.17 |
| 3 | Multilayer Perceptron | 0.359 | 0.4601 | 47.98 | 47.00 |
| 4 | *Random Forest* | *0.1253* | *0.2589* | *16.75* | *26.44* |
| 5 | Random Tree | 0.1 | 0.3162 | 13.36 | 32.30 |

## 7.3 Discussion of Findings

### 7.3.1 Objective 1: Applied Sciences in Marvel

The applied science fields depicted in the MCU movies can be grouped into four Super Domains: i) Technology, ii) Natural Sciences, ii) Fictional Technology, and iv) Sciences. The following sub domains are further explained and portrayed in the movies Technology, AI, Robotics, Magic, Ancient Technology, Cosmic Technology, Multiverse Technology, Energy, Physics, Engineering. Technology and physics are the most frequently shown applied science areas in MCU films. This is hardly surprising given the importance of these subjects in the development of advanced technologies and weapons utilized by superheroes and villains. Among the specific examples of applied science technology featured in MCU films are:

1. AI: Tony Stark's JARVIS and FRIDAY AI assistants, Ultron, Vision
2. Robotics: Iron Man's suits, Spider-Man's suit, War Machine's armor
3. Fictional Technology: The Tesseract, the Infinity Gauntlet, Captain America's shield
4. Magic: Doctor Strange's spells, the Cloak of Levitation, the Eye of Agamotto
5. Ancient Technology: The Wakandans' Vibranium technology
6. Cosmic Technology: The Asgardians' Bifrost Bridge, the Guardians of the Galaxy's Milano spacecraft
7. Multiverse Technology: The TVA's TemPads
8. Energy: The Stark Tower's arc reactor, the Avengers Compound's solar panels
9. Physics: The Hulk's gamma ray-powered transformations, Iron Man's repulsor rays, Thor's lightning powers
10. Engineering: The Helicarrier, the Quinjet, Wakanda's Maglev train.

It should be noted that the MCU films are works of fiction, and the science and technology represented in them are not completely realistic. The MCU films, on the other hand, do an excellent job of demonstrating how applied science can be utilized to solve issues and enhance people's lives.

Aside from the applied science subjects mentioned above, the MCU films show a variety of additional scientific topics, including quantum mechanics, the multiverse, and time travel. These ideas are frequently employed to develop suspenseful and interesting tales. The MCU films, on the other hand, investigate the ethical ramifications of this technology, such as the possible hazards of artificial intelligence and the consequences of changing the past. Overall, the MCU films provide an excellent overview of a diverse range of applied science topics. The films demonstrate how science may be utilized to develop both useful and destructive technologies. They also investigate the ethical consequences of scientific breakthroughs. MCU films can be an excellent instructional resource for students of science and engineering.

### 7.3.2 Objective 2: Nexus of Technology Representation and Box Office Success

According to the statistics and data analytics findings, there is a substantial association between the portrayal of technology in MCU films and their box office performance. Movies having a higher percentage of technical content are more profitable.

This correlation can be seen in the data in Tables 9 and 10. Table 9 indicates that if the Tech Content Count is 0, no movie falls into Class 1 (maximum profitability), but all the highly profitable movies fall into Class 1. Similarly, Table 10 reveals that a considerable proportion of films (7) are classified as Class 4 (lowest profitability) and have a Tech Content Count of 0. Table 10 shows the correlation analysis results, which reveal a substantial association between the Tech Content Count and the Profit Percentage. The Random Forest technique, which achieved the highest correlation coefficient (0.964), indicates that the two variables have a significant non-linear relationship. The regression analysis findings, as shown in Table 11, reveal that the Random Forest method likewise had the lowest error margin, implying that it is the most accurate model for forecasting the Profit Percentage of an MCU movie based on its Tech Content Count.

One explanation for this correlation is because spectators are drawn to the spectacle of sophisticated technology and the dramatic action sequences it allows. The MCU movies, for example, have a variety of remarkable technological equipment, such as Iron Man's armor, Captain America's shield, and Thor's hammer. These devices enable superheroes to accomplish great feats such as flying, fighting supervillains, and saving the world. Another argument for the link could be that the MCU films use technology to explore complicated themes and ideas. The Iron Man films, for example, investigate the ethical implications of artificial intelligence, whereas the captain America films investigate the relationship between technology and freedom. This enables the MCU films to appeal to a diverse audience, including those interested in science and technology.

## 8. Limitations and Future Directions
### 8.1 Limitations of the study

This research has focused on the Technology representation influence on Box Office performance. Whereas three other domains (Science, Natural Science and Fictional Technology) are also identified from the content analysis, the limitation of our study is that its only focus and targeted sector is technology on MCU movies. Whereas Science Fiction cinema has a high rate of movie release with respect to other genres.

### 8.2 Future research directions

Despite the pattern is identified that technology-oriented movies have prominent ratio of momentous success on Box Office, still the other domains' study is also required to get the insight on Box Office rate of Science Fiction movies, moreover this study can further be expanded on other movie franchise for the bigger exposure like DCU and Star Wars. So, more research on the role of applied science in a movie's success will make the domain a multidisciplinary zone combining entertainment, academics, literature, and media industries together on a broader scale.

## 9. Conclusion

The analysis of the applied science subjects depicted in the MCU movies, as well as the relationship between the portrayal of technology and box office performance, indicates that technology plays a key part in the MCU films' appeal. Technology, magic, ancient technology, cosmic technology, multiverse technology, energy, physics, and engineering are all portrayed in MCU movies. Technology and physics are the most frequently shown applied science areas in MCU movies. The representation of technology in MCU films and their box office success are strongly linked. Movies having a higher percentage of technical content are more profitable. This is due to the spectacle of sophisticated technology and the exciting action sequences that it allows. Furthermore, the MCU movies use technology to explore complicated topics and ideas, allowing them to appeal to a diverse audience.

Overall, the study's findings indicate that the portrayal of technology is one among many variables that contribute to the success of MCU films. Audiences are drawn to the sight of sophisticated technology, the exciting action sequences it permits, and the profound themes and ideas that technology explores. But it is crucial to keep in mind that technology is not a panacea. To use technology effectively, careful preparation and execution are required.

**Statements and Declarations**

**Competing interest -** All the Authors state no conflict-of-interest whether financial or non-financial.